\documentclass{svmult}
\usepackage{amsbsy,amssymb,amscd,amsfonts,latexsym,amstext,delarray,
amsmath,epsfig, graphicx}

\newtheorem{thm}{Theorem}[section]

\newtheorem{lem}[thm]{Lemma}

\def\R{{\mathbb R}}
\def\H{{\mathbb H}}
\def\Z{{\mathbb Z}}
\def\Q{{\mathbb Q}}
\def\C{{\mathbb C}}

\def\P{{\mathbb P}}

\def\F{{\mathbb F}}

\def\PGL{{\rm PGL}}

\def\cA{{\mathcal A}}

\def\cF{{\mathcal F}}
\def\cL{{\mathcal L}}

\newcommand{\ie}{{\it i.e.\/}\ }

\newcommand{\cf}{{\it cf.\/}\ }

\begin{document}

\title*{Modular curves, ${\rm C}^*$-algebras, and chaotic cosmology}
\author{Matilde Marcolli}
\institute{Max--Planck Institut f\"ur
Mathematik, Bonn, Germany; marcolli\@@mpim-bonn.mpg.de}
\maketitle

We make some brief remarks on the relation of the mixmaster universe
model of chaotic cosmology to the geometry of modular curves
and to noncommutative geometry. 
We show that the full dynamics of the mixmaster universe is
equivalent to the geodesic flow on the modular curve
$X_{\Gamma_0(2)}$. We then consider a special class of solutions,
with bounded number of cycles in each Kasner era, and describe their
dynamical properties (invariant density, Lyapunov exponent, topological
pressure). We relate these properties to the noncommutative geometry
of a moduli space of such solutions, which is given by a
Cuntz--Krieger ${\rm C}^*$-algebra.

\section{Modular curves}

Let $G$ be a finite index subgroup of $\Gamma =\PGL(2,\Z)$, and
let $X_G$ denote the quotient $X_G = G \backslash \H^2$, where
$\H^2$ is the 2-dimensional real hyperbolic plane, namely the
upper half plane $\{ z\in \C : \Im  z > 0 \}$ with the metric
$ds^2=|dz|^2/(\Im z)^2$. Equivalently, we identify $\H^2$ with the
Poincar\'e disk $\{ z : | z |<1 \}$ with the metric $ds^2 = 4
|dz|^2/(1-|z|^2)^2$.

Let $\P$ denote the coset space $\P = \Gamma/G$. We can write the
quotient $X_G$ equivalently as $X_G = \Gamma \backslash (\H^2
\times \P)$. The quotient space $X_G$ has the structure of a
non-compact Riemann surface, which can be compactified by adding
cusp points at infinity: 
\begin{equation}\label{alg-compact}
\bar X_G= G \backslash ( \H^2 \cup \P^1(\Q)) \simeq \Gamma
\backslash \left( ( \H^2 \cup \P^1(\Q)) \times \P \right).
\end{equation} 
In particular, we consider the congruence subgroups
$G=\Gamma_0(p)$, with $p$ a prime, given by matrices
$$ g=\left(\begin{array}{cc} a&b\\c&d \end{array}\right) $$
satisfying $c \equiv 0 \mod p$. In fact, for our purposes, we are
especially interested in the case $p=2$.

\subsection{Shift operator and dynamics}

If we consider the boundary $\P^1(\R)$ of $\H^2$, the arguments
given in \cite{ManMar} and \cite{Mar} show that the quotient
$\Gamma \backslash (\P^1(\R)\times \P)$, better interpreted as a
``noncommutative space'', gives rise to a compactification of the
modular curve $X_G$ with a structure richer than the ordinary
algebro-geometric compactification by cusp points.

In \cite{ManMar} this was described in terms of the following
dynamical system, generalizing the classical Gauss shift of the
continued fraction expansion: 
$$ T: [0,1] \times \P \to [0,1] \times \P $$
\begin{equation}\label{T}
 T(x,t) = \left( \frac{1}{x} - \left[
\frac{1}{x} \right], \left(\begin{array}{cc} -[1/x] & 1 \\
1 & 0 \end{array}\right)\cdot t \right).
\end{equation} 
In fact, the quotient space of $\P^1(\R)\times \P$ by the
$\PGL(2,\Z)$ can be identified with the space of orbits of the
dynamical system $T$ on $[0,1]\times \P$.

We use the notation $x=[k_1,k_2,\ldots, k_n, \ldots ]$ for the
continued fraction expansion of the point $x\in [0,1]$, and we
denote by $p_n(x)/q_n(x)$ the convergents of the continued
fraction, with $p_n(x)$ and $q_n(x)$ the successive
numerators and denominators. It is then easy to verify that the
shift $T$ acts on $x$ by shifting the continued fraction
expansion, $T [k_1,k_2,\ldots,
k_n,\ldots]=[k_2,k_3,\ldots,k_{n+1},\ldots]$. The element
$g_n(x)^{-1}$, with
$$
g_n(x)=\left(\begin{array}{cc}
p_{k-1}(x ) & p_k(x )\\
q_{k-1}(x ) & q_k(x ) \end{array} \right) \in \Gamma,
$$
acts on $[0,1]\times \P$ as $T^n$.

The Lyapunov exponent
$$ \lambda(x) := \lim_{n\to \infty}
\frac{1}{n} \log | (T^n)^\prime (x) |
$$
$$ =2 \lim_{n\to \infty} \frac{1}{n} \log q_n(x) $$
measures the exponential rate of divergence of nearby orbits,
hence it provides a measure of how chaotic the dynamics is.

For the shift of the continued fraction expansion, the results of
\cite{PoWei} show that $\lambda(x)=\pi^2 /(6\log 2)=\lambda_0$
almost everywhere with respect to the Lebesgue measure on $[0,1]$
and counting measure on $\P$. On the other hand, the Lyapunov
exponent takes all values $\lambda(x)\in [\lambda_0,\infty)$.
The unit interval correspondingly splits as a union of
$T$-invariant level sets of $\lambda$ (Lyapunov spectrum) of
varying Hausdorff dimension, plus an exceptional set where the
limit defining $\lambda$ does not converge.

\section{Mixmaster universe}

An important problem in cosmology is understanding how anisotropy
in the early universe affects the long time evolution of
space-time. This problem is relevant to the study of the beginning
of galaxy formation and in relating the anisotropy of the
background radiation to the appearance of the universe today.

We follow \cite{Barrow} (\cf also \cite{MTW}) for a brief summary of
anisotropic and chaotic cosmology. The simplest significant
cosmological model that presents strong anisotropic properties is
given by the Kasner metric 
\begin{equation}\label{Kasner}
ds^2 = -dt^2 + t^{2p_1} dx^2 + t^{2p_2} dy^2 + t^{2p_3} dz^2,
\end{equation} 
where the exponents $p_i$ are constants satisfying $\sum
p_i=1=\sum_i p_i^2$. Notice that, for $p_i=d\log g_{ii}/d\log g$,
the first constraint $\sum_i p_i=1$ is just the condition that
$\log g_{ij}= 2\alpha \delta_{ij} + \beta_{ij}$ for a traceless
$\beta$, while the second constraint $\sum_i p_i^2=1$ amounts to
the condition that, in the Einstein equations written in terms of
$\alpha$ and $\beta_{ij}$, 
$$ \left(\frac{d\alpha}{dt}\right)^2= \frac{8\pi}{3} \left( T^{00}
+ \frac{1}{16\pi} \left( \frac{d\beta_{ij}}{dt}\right)^2 \right)
$$
$$ e^{-3\alpha} \frac{d}{dt} \left(
e^{3\alpha}\frac{d\beta_{ij}}{dt} \right) = 8\pi \left( T_{ij}
-\frac{1}{3} \delta_{ij} T_{kk} \right),
$$ 
the term $T^{00}$ is negligible with respect to the term
$(d\beta_{ij}/dt)^2/16\pi$, which is the ``effective energy
density'' of the anisotropic motion of empty space, contributing
together with a matter term to the Hubble constant.

Around 1970, Belinsky, Khalatnikov, and Lifshitz introduced a
cosmological model ({\em mixmaster universe}) where they allowed
the exponents $p_i$ of the Kasner metric to depend on a parameter
$u$,
\begin{equation}\label{u-param}
\begin{array}{l}
p_1=\frac{-u}{1+u+u^2} \\[3mm]
p_2=\frac{1+u}{1+u+u^2}\\[3mm]
p_3=\frac{u(1+u)}{1+u+u^2}
\end{array}
\end{equation}
Since for fixed $u$ the model is given by a Kasner space-time, the
behavior of this universe can be approximated for certain large
intervals of time by a Kasner metric. In fact, the evolution is
divided into Kasner eras and each era into cycles. During each era
the mixmaster universe goes through a volume compression. Instead
of resulting in a collapse, as with the Kasner metric, high
negative curvature develops resulting in a bounce (transition to a
new era) which starts again a behavior approximated by a Kasner
metric, but with a different value of the parameter $u$. Within
each era, most of the volume compression is due to the scale
factors along one of the space axes, while the other scale factors
alternate between phases of contraction and expansion. These
alternating phases separate cycles within each era.

More precisely, we are considering a metric
\begin{equation}\label{mixmaster}
ds^2 =-dt^2 + a(t) dx^2 + b(t) dy^2 + c(t) dz^2,
\end{equation}
generalizing the Kasner metric \eqref{Kasner}. We still require
that \eqref{mixmaster} admits $SO(3)$ symmetry on the space-like
hypersurfaces, and that it presents a singularity at $t\to 0$. In
terms of logarithmic time $d\Omega= -\frac{dt}{abc}$, the {\em
mixmaster universe} model of Belinsky, Khalatnikov, and Lifshitz
admits a discretization with the following properties:

\medskip

\noindent{\bf 1.} The time evolution is divided in Kasner eras
$[\Omega_n, \Omega_{n+1}]$, for $n\in \Z$. At the beginning of
each era we have a corresponding discrete value of the parameter
$u_n > 1$ in \eqref{u-param}.

\noindent{\bf 2.} Each era, where the parameter $u$ decreases with
growing $\Omega$, can be subdivided in cycles corresponding to the
discrete steps $u_n$, $u_n -1$, $u_n -2$, etc. A change $u\to u-1$
corresponds, after acting with the permutation $(12)(3)$ on the
space coordinates, to changing $u$ to $-u$, hence replacing
contraction with expansion and conversely. Within each cycle the
space-time metric is approximated by the Kasner metric
\eqref{Kasner} with the exponents $p_i$ in \eqref{u-param} with a
fixed value of $u=u_n -k >1$.

\noindent{\bf 3.} An era ends when, after a number of cycles, the
parameter $u_n$ falls in the range $0< u_n < 1$. Then the bouncing
is given by the transition $u\to 1/u$ which starts a new series of
cycles with new Kasner parameters and a permutation $(1)(23)$ of
the space axis, in order to have again $p_1< p_2< p_3$ as in
\eqref{u-param}.

\bigskip

Thus, the transition formula relating the values $u_n$ and
$u_{n+1}$ of two successive Kasner eras is
$$ u_{n+1} = \frac{1}{u_n - [u_n]}, $$
which is exactly the shift of the continued fraction expansion,
$Tx=1/x -[1/x]$, with $x_{n+1}=T x_n$ and $u_n = 1/x_n$.

\section{Geodesics and universes}

The previous observation is the key to a geometric description of
solutions of the mixmaster universe in terms of geodesics on a
modular curve (Manin--Marcolli \cite{ManMar}):

\begin{thm}\label{thm-geod}
Consider the modular curve $X_{\Gamma_0(2)}$. Each infinite
geodesic $\gamma$ on $X_{\Gamma_0(2)}$ not ending at cusps
determines a mixmaster universe.
\end{thm}

\begin{proof} An infinite geodesic on
$X_{\Gamma_0(2)}$ is the image under the quotient map
$$
\pi_\Gamma : \H^2 \times \P \to \Gamma\backslash (\H^2\times \P)
\cong X_G,
$$
where $\Gamma =\PGL(2,\Z)$, $G=\Gamma_0(2)$, and $\P=\Gamma/G\cong
\P^1(\F_2)=\{ 0,1,\infty \}$, of an infinite geodesic on
$\H^2\times \P$ with ends on $\P^1(\R)\times\P$. We consider the
elements of $\P^1(\F_2)$ as labels assigned to the three space
axes, according to the identification
\begin{equation}\label{axes}
\begin{array}{lll}
0=[0:1]& \mapsto & z \\
\infty=[1:0] & \mapsto & y \\
1=[1:1] & \mapsto & x.
\end{array}
\end{equation}

Any geodesic not ending at cups can be coded in terms of data
$(\omega^-,\omega^+, s)$, where $(\omega^{\pm},s)$ are the
endpoints in $\P^1(\R)\times \{ s \}$, $s\in \P$, with
$\omega^-\in (-\infty,-1]$ and $\omega^+\in [0,1]$. In terms of
the continued fraction expansion, we can write
$$ \begin{array}{ll}\omega^+ & = [k_0, k_1,
\ldots k_r, k_{r+1}, \ldots] \\
 \omega^- & =[k_{-1};k_{-2},\ldots, k_{-n}, k_{-n-1}, \ldots].
\end{array} $$
The shift acts on these data by 
$$ T (\omega^+,s)= \left( \frac{1}{\omega^+} -\left[
\frac{1}{\omega^+} \right], \left(\begin{array}{cc}
-[1/\omega^+]&1 \\ 1&0 \end{array}\right) \cdot s\right) $$
$$ T(\omega^-,s)=\left( \frac{1}{\omega^- +[1/\omega^+]},
\left(\begin{array}{cc} -[1/\omega^+]&1
\\ 1&0 \end{array}\right) \cdot s\right). $$ 
Geodesics on $X_{\Gamma_0(2)}$ can be identified with the orbits
of $T$ on the set of data $(\omega,s)$.

The data $(\omega,s)$ determine a mixmaster universe, with the
$k_n=[u_n]=[1/x_n]$ in the Kasner eras, and with the transition
between subsequent Kasner eras given by $x_{n+1}=Tx_n \in [0,1]$
and by the permutation of axes induced by the transformation
$$ \left( \begin{array}{cc} -k_n &1 \\ 1 & 0 \end{array}\right) $$
acting on $\P^1(\F_2)$. It is easy to verify that, in fact, this
acts as the permutation $0\mapsto \infty$, $1\mapsto 1$,
$\infty\mapsto 0$, if $k_n$ is even, and $0\mapsto \infty$,
$1\mapsto 0$, $\infty \mapsto 1$ if $k_n$ is odd, that is, after
the identification \eqref{axes}, as the permutation $(1)(23)$ of
the space axes $(x,y,z)$, if $k_n$ is even, or as the product of
the permutations $(12)(3)$ and $(1)(23)$ if $k_n$ is odd. This is
precisely what is obtained in the mixmaster universe model by the
repeated series of cycles within a Kasner era followed by the
transition to the next era.

Data $(\omega,s)$ and $T^m (\omega,s)$, $m\in \Z$, determine the
same solution up to a different choice of the initial time.

There is an additional time-symmetry in this model of the evolution of
mixmaster universes (\cf \cite{Barrow}). In fact, there is an
additional parameter $\delta_n$ in the system, which measures the initial
amplitude of each cycle. It is shown in \cite{Barrow} that this is
governed by the evolution of a parameter 
$$ v_n = \frac{\delta_{n+1} (1+u_n)}{1-\delta_{n+1}} $$
which is subject to the transformation across cycles
$v_{n+1} = [u_n] + v_n^{-1}$. 
By setting $y_n =v_n^{-1}$ we obtain
$$ y_{n+1} = \frac{1}{\left( y_n + \left[ 1/x_n \right] \right)},
$$ 
hence we see that we can interpret the
evolution determined by the data $(\omega^+,\omega^-,s)$ with the
shift $T$ either as giving the complete evolution of the 
$u$-parameter towards and away from the cosmological singularity, or
as giving the simultaneous evolution of the two parameters $(u,v)$
while approaching the cosmological singularity. 

This in turn determines the complete evolution of the parameters
$(u,\delta,\Omega)$, where $\Omega_n$ is the starting time of each
cycle. For the explicit recursion $\Omega_{n+1}=\Omega_{n+1}
(\Omega_n,x_n,y_n)$ see \cite{Barrow}.

\end{proof}

Notice that, unlike the description of the full mixmaster dynamics
given, for instance, in \cite{Barrow}, we {\em include} the
alternation of the space axes at the end of cycles and eras as part of
the dynamics, which is precisely what determines the choice of the
congruence subgroup. This also
introduces a slight modification of some of the invariants
associated to the mixmaster dynamics. For instance, it is proved in
\cite{ManMar} that there is a unique $T$--invariant measure on
$[0,1]\times \P$, which is given by the 
Gauss density on $[0,1]$ and the counting measure $\delta$ on
$\P$:
\begin{equation}\label{inv-meas}
 d\mu(x,s)= \frac{\delta(s)\, dx}{3\log(2)\, (1+x)}, 
\end{equation}
which reduces to the Gauss density for the
shift of the continued fraction on $[0,1]$ when integrated in the $\P$
direction. In particular, as observed 
in \cite{ManMar}, the form \eqref{inv-meas} of the invariant measure
implies that the alternation of the space axes 
is uniform over the time evolution, namely the three axes provide 
the scale factor responsible for volume compression with equal
frequencies.

\section{Controlled pulse universes}

The interpretation of solutions in terms of geodesics provides a
natural way to single out and study certain special classes of
solutions on the basis of their geometric properties. Physically, such
special classes of solutions exhibit different behaviors 
approaching the cosmological singularity.

For instance, the data
$(\omega^+,s)$ corresponding to an eventually periodic sequence
$k_0 k_1 \ldots k_m \ldots$
of some period $\overline{a_0\ldots a_\ell}$ correspond to those
geodesics on $X_{\Gamma_0(2)}$ that asymptotically wind around the
closed geodesic identified with the doubly infinite sequence
$\ldots a_0\ldots a_\ell a_0\ldots a_\ell \ldots$.
Physically, these universes exhibit a pattern of cycles that
recurs periodically after a finite number of Kasner eras.

In the following we concentrate on another special class of solutions,
which we call {\em controlled pulse universes}. These are the mixmaster
universes for which there is a fixed upper bound $N$ to the number of
cycles in each Kasner era. 

In terms of the continued fraction description, these solutions
correspond to data $(\omega^+,s)$ with $\omega^+$ in the Hensley
Cantor set $E_N\subset [0,1]$. The set $E_N$ is given by all points in
$[0,1]$ with all digits in the continued fraction expansion bounded by
$N$ (\cf \cite{Hen2}). In more geometric terms, one considers geodesics 
on the modular curve $X_{\Gamma_0(2)}$ that wander only a finite
distance into the cusps.  

\subsection{Dynamical properties}

It is well known that a very effective technique for the study of
dynamical properties such as topological pressure and invariant
densities is given by transfer operator methods. These have
already been applied successfully to the case of the mixmaster
universe, \cf \cite{Mayer}.  
In our setting, for the full mixmaster dynamics that includes
alternation of space axes, the Perron-Frobenius operator for the shift
\eqref{T} is of the form 
$$
({\mathcal L}_{\beta} f)(x,s)  = \sum_{k=1}^\infty
\frac{1}{(x+k)^{\beta}} f \left( \frac{1}{x +k},
\left(\begin{array}{cc} 0 & 1\\ 1 & k \end{array} \right) \cdot s
\right).  
$$ 
This yields the density of the invariant measure \eqref{inv-meas}
satisfying ${\mathcal L}_{2} f= f$. The top eigenvalue $\eta_\beta$
of ${\mathcal L}_{\beta}$ is related to the topological pressure by
$\eta_\beta=\exp( P(\beta) )$. This can be estimated numerically,
using the technique of \cite{Babenko} and the integral kernel operator
representation of \S 1.3 of \cite{ManMar}.

We now restrict our attention to the case of controlled pulse
universes. Since these form sets of measure zero in the measure
\eqref{inv-meas}, they support exceptional values of such dynamical
invariants as Lyapunov exponent, topological pressure, entropy. In
fact, for a fixed bound $N$ on the number of cycles per era, we are
considering the dynamical system \eqref{T}
$$ T: E_N \times \P \to E_N \times \P. $$
For this map, the Perron-Frobenius operator is of the form
\begin{equation}\label{PFN}
({\mathcal L}_{\beta,N} f)(x,s)  = \sum_{k=1}^N
\frac{1}{(x+k)^{\beta}} f \left( \frac{1}{x +k},
\left(\begin{array}{cc} 0 & 1\\ 1 & k \end{array} \right) \cdot s
\right).  
\end{equation} 
It is proved in \cite{Mar} that this operator still
has a unique invariant measure $\mu_N$, whose density satisfies
${\mathcal L}_{2\dim_H(E_N),N} f = f$, with 
$$ \dim_H(E_N) = 1- \frac{6}{\pi^2 N} - \frac{72 \log N}{
\pi^4 N^2} + O(1/N^2) $$ 
the Hausdorff dimension of the Cantor set $E_N$.
Moreover, the top eigenvalue $\eta_\beta$
of ${\mathcal L}_{\beta,N}$ is related to the 
Lyapunov exponent by 
$$ \lambda(x) = 2 \frac{d}{d\beta} \eta_\beta|_{\beta =2\dim_H(E_N)}, $$
for $\mu_N$-almost all $x\in E_N$, \cf \cite{Mar}.

\section{Non-commutative spaces}

A consequence of this characterization of the time evolution in terms
of the dynamical system \eqref{T} is that 
we can study global properties of suitable {\em moduli spaces} of
mixmaster universes. For instance, the moduli space for time
evolutions of the $u$-parameter approaching the cosmological
singularity as $\Omega \to \infty$ is given by the quotient of 
$[0,1]\times \P$ by the action of the shift $T$. Similarly, the moduli
spaces that correspond to controlled pulse universes are the quotients of 
$E_N \times \P$ by the action of the shift $T$. It is easy to see
that such quotients are not well behaved as a topological spaces,
which makes it difficult to study their global properties in the
context of classical geometry. However, non-commutative geometry in
the sense of Connes \cite{Connes} provides the correct framework for
the study of such spaces. The occurrence in physics of non-commutative
spaces as moduli spaces is not a new phenomenon. A well known example
is the moduli space of Penrose tilings (\cf \cite{Connes}), which
plays an important role in the mathematical theory of quasi-crystals.

We consider here only the case of controlled pulse universes. In this case,
the dynamical system $T$ is a subshift of finite type which can be
described by the Markov partition    
$$ {\mathcal A}_N = \{ ((k,t),(\ell,s)) | U_{k,t} \subset
T(U_{\ell,s}) \}, 
$$ 
for $k,\ell \in \{ 1, \ldots, N \}$, and $s,t\in \P$, with sets
$U_{k,t} = U_k \times \{ t \}$, where $U_k \subset E_N$ are the clopen
subsets where the local inverses of $T$ are defined,
$$ U_k = \left[\frac{1}{k+1},\frac{1}{k}\right]\cap E_N. $$
This Markov partition determines a matrix $A_N$, with entries
$(A_N)_{kt,\ell s} =1$ if $U_{k,t} \subset T(U_{\ell,s})$ and zero
otherwise. 

\begin{lem}\label{ANlem}
The $3\times 3$ submatrices $A_{k\ell}=( A_{(k,t),(\ell,s)} )_{s,t
\in\P}$ of the matrix $A_N$ are of the form
$$ A_{k\ell}=\left\{\begin{array}{ll} M_1=\left(\begin{array}{ccc} 
0&0&1\\ 0&1&0\\ 1&0&0 \end{array}\right) & \ell =2m \\[6mm]
M_2=\left(\begin{array}{ccc} 
0&0&1\\ 1&0&0\\ 0&1&0 \end{array}\right) & \ell =2m +1
\end{array}\right. $$
The matrix $A_N$ is irreducible and aperiodic.
\end{lem}

\begin{proof} The condition $U_{k,t} \subset
T(U_{\ell,s})$ is equivalently written as the condition that
$$ \left(\begin{array}{cc} 0 & 1 \\ 1 & \ell \end{array} \right) \cdot s
=t . $$
We then proceed as in the proof of Theorem \ref{thm-geod} and notice
that the transformation 
$$ \left(\begin{array}{cc} 0 & 1 \\ 1 & \ell \end{array} \right) $$
acts on $\P^1(\F_2)$, under the identification \eqref{axes}, as the
permutation $0\mapsto \infty$, $1\mapsto 1$,
$\infty\mapsto 0$, when $\ell$ is even, and $\infty \mapsto 0$,
$0\mapsto 1$, $1\mapsto \infty$ if $\ell$ is odd.

Irreducibility of $A_N$ means that the corresponding directed graph is
strongly connected, namely any two vertices are connected by an
oriented path of edges. Since the matrix $A_N$ has the form
$$ A_N = \left( \begin{array}{ccccc} M_1 & M_1 & M_1 & \cdots & M_1 \\
M_2 & M_2 & M_2 & \cdots & M_2 \\
M_1 & M_1 & M_1 & \cdots & M_1 \\
\vdots & & & \cdots & \vdots \end{array} \right), $$
irreducibility follows from the irreducibility of $A_2$, which
corresponds to the directed graph illustrated in the Figure.
\begin{figure}
\begin{center}
\epsfig{file=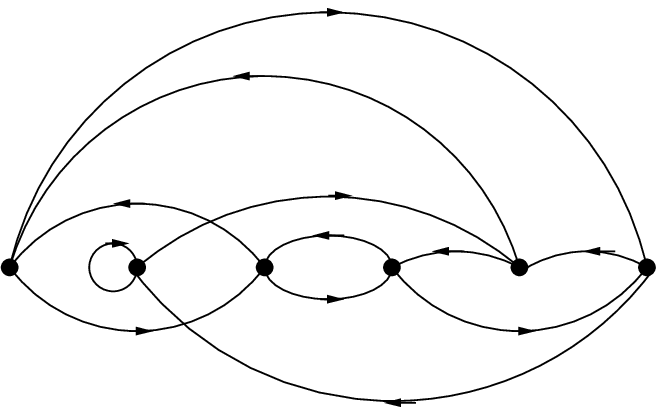} 
\end{center}
\end{figure}
This graph also shows that the matrix $A_N$ is aperiodic. In fact, the
period is defined as the gcd of the lengths of the closed directed
paths and the matrix is called aperiodic when the period is equal to one.

\end{proof}

As a non-commutative space associated to the Markov partition we
consider the Cuntz--Krieger ${\rm C}^*$-algebra ${\mathcal O}_{A_N}$
(\cf \cite{CuKri}), which is the universal ${\rm C}^*$-algebra
generated by partial isometries $S_{kt}$ satisfying the relations 
$$ \sum_{(k,t)}S_{kt}S_{kt}^* = 1, $$
$$ S_{\ell s}^* S_{\ell s} =\sum_{(k,t)} A_{(k,t),(\ell,s)}
S_{kt}S_{kt}^*. $$
 
It is a well known fact that the structure of this ${\rm
C}^*$-algebra reflect properties of the dynamics of the shift $T$. 
For example, one can recover the Bowen--Franks invariant  
of the dynamical system $T$ from the $K$-theory of a Cuntz--Krieger
${\rm C}^*$-algebra ${\mathcal O}_{A}$ (\cf \cite{CuKri}).
Moreover, in our set of examples, information on the dynamical 
properties of the shift $T$ and the Perron--Frobenius
operator \eqref{PFN} can be derived from the KMS states for the
$C^*$-algebras ${\mathcal O}_{A_N}$ with respect to suitable time
evolutions. 

\subsection{Time evolution and KMS states}

Recall that a state $\varphi$ on a unital $C^*$-algebra $\cA$ is a
continuous linear functional $\varphi: \cA \to \C$ satisfying
$\varphi(a^*a)\geq 0$ and $\varphi(1)=1$.  
Let $\sigma_t$ be an action of $\R$ on $\cA$ by automorphisms.   
A state $\varphi$ satisfies the KMS condition at inverse temperature
$\beta$ if for any $a,b\in \cA$ there exists a bounded
holomorphic function $F_{a,b}$ continuous on $0\leq Im(z) \leq \beta$,
such that, for all $t\in\R$,
\begin{equation}\label{KMS}
F_{a,b}(t)=\varphi(a\,\sigma_t(b))\ \ \text{ and } \ \
F_{a,b}(t+i\beta) = \varphi(\sigma_t(b\,)a) 
\end{equation}
Equivalently, the KMS condition \eqref{KMS} is expressed as the
relation  
\begin{equation}\label{KMS2}
\varphi( \sigma_t(a)\, b ) = \varphi( b\, \sigma_{t+i\beta}(a)).
\end{equation}

Consider now a Cuntz--Krieger ${\rm C}^*$-algebra ${\mathcal O}_{A}$. 
Any element in the $*$-algebra generated algebraically by the $S_i$ 
can be written as a linear combination of monomials of the form
$S_\mu S_\nu^*$, for multi-indices $\mu=(i_1,\ldots i_{|\mu|})$ and
$\nu=(j_1,\ldots, j_{|\nu|})$. 
The subalgebra $\cF_A$ is the AF-algebra generated by the elements of
the form $S_\mu S_\nu^*$, for 
$|\mu |=|\nu |$. It is filtered by finite dimensional subalgebras
$\cF_k$, for $k\geq 0$, generated by the matrix units $E^i_{\mu,\nu}=
S_\mu P_i S_\nu^*$, with $|\mu |=|\nu |=k$, where $P_i=S_i S_i^*$ are
the range projections of the isometries $S_i$.
The commutative algebra of functions on the Cantor set $\Lambda_A$ of
the subshift of finite type $(\Lambda_A, T)$ associated to the
Cuntz--Krieger algebra, is a maximal abelian subalgebra of $\cF_A$,
identified with the elements of the form $S_\mu S_\mu^*$ (\cf
\cite{CuKri}).  

In our case, one can consider a time evolution
on ${\mathcal O}_{A_N}$ of the type considered in
\cite{KerrPinz}, with $\sigma_t^{u,h}(S_k)=\exp(it(u-h))\, S_k$, for 
$u = \log \eta_\beta = P(\beta)$, the topological pressure, \ie the
top eigenvalue of \eqref{PFN}, and $h(x)= -\beta/2\, \log | T'(x)
|$. Thus, for all $t\in \R$, the function $\exp(-it\, h)$ acts on the
elements of ${\mathcal O}_{A_N}$ by multiplication by an element in
$C(E_N\times \P)$.  

Then a KMS$_1$ state $\varphi$ for this time evolution satisfies the
relation (\cf \cite{KerrPinz} Lemma 7.3)
$$ \sum_k \varphi (S^*_k \, e^h a \, S_k) = e^u \, \varphi (a), $$
for $a\in {\mathcal O}_{A_N}$. For all $a=f\in C(E_N\times \P)$, we
have $\sum_k S^*_k \, e^h f \, S_k= \cL_h (f)$,
where the Ruelle transfer operator 
$$ \cL_h (f)(x,s) =\sum_{(y,r)\in T^{-1}(x,s)} \exp(h(y)) \, f(y,r) $$
is in fact the Perron--Frobenius operator \eqref{PFN}. In particular,
the KMS condition implies that the state $\varphi$ restricts to 
$C(E_N\times \P)$ to a probability measure $\mu$ satisfying $\cL^*_h
\mu =e^u \mu$. For $u=\log \eta_\beta$, the existence and uniqueness
of such measure can be derived from the properties of the operator
\eqref{PFN}, along the lines of \cite{Mayer2} \cite{ManMar}. 

\begin{thm}
For $\beta < 2 \log r(A_N)/ \log (N+1)$, there exists a unique KMS$_1$
state for the time evolution $\sigma_t^{P(\beta),-\beta/2\, \log |
T'|}$ on the algebra ${\mathcal O}_{A_N}$. This restricts to the
subalgebra $C(E_N\times \P)$ to $f \mapsto \int f\, d\mu$ with the
probability measure satisfying 
$\cL^*_\beta \mu =\eta_\beta \mu$, for $\cL_\beta$ the
Perron--Frobenius operator \eqref{PFN}.
\end{thm}

\begin{proof} Since by Lemma \ref{ANlem} the matrix $A_N$ is
irreducible and aperiodic, by Proposition 7.6 of \cite{KerrPinz},
there is a surjective map of the set of KMS 
states to the set of probability measures satisfying $\cL^*_h
\mu =e^u \mu$. Uniqueness follows from Lemma 7.5
and Theorem 7.8 of \cite{KerrPinz}, by showing that the estimate
$var_0(h)< r(A_N)$ holds, where $r(A_N)$ is the spectral radius
of the matrix $A_N$ and $var_0(h)= \max h - \min h$ on $E_N\times
\P$. This provides the range of values of $\beta$ specified above,
since on $E_N\times \P$ we have $var_0(-\beta/2 \log |T'|)=
\log(N+1)^{\beta/2}$.  The KMS state $\varphi$ is obtained as in
\cite{KerrPinz}, by defining inductively compatible states
$$ \varphi_k (a) = e^{-u} \sum_j \varphi_{k-1} (S_j^* e^{h/2} a
e^{h/2} S_j) \ \ \ \text{ for } \ \ a\in \cF_k. $$ 

\end{proof}


\begin{thebibliography}{99}

\bibitem{Babenko} K.~I.~Babenko, {\it On a problem of Gauss}.
Dokl. Akad. Nauk SSSR, Tom 238 (1978) No. 5, 1021--1024.

\bibitem{Barrow} J.~D.~Barrow. {\it Chaotic behaviour and the Einstein
equations.} In: Classical General Relativity, eds. W.~Bonnor et
al., Cambridge Univ. Press, Cambridge, 1984, 25--41.

\bibitem{BKL} V.~Belinskii, I.~M.~Khalatnikov, E.~M.~Lifshitz. {\it
Oscillatory approach to singular point in Relativistic cosmology.}
Adv.~Phys. 19 (1970), 525-551.

\bibitem{Connes} A.~Connes, {\it Noncommutative Geometry}, Academic
Press, 1994. 

\bibitem{CuKri} J.~Cuntz, W.~Krieger, {\it A class of $C^*$--algebras
and topological Markov chains}, Invent. Math. 56 (1980) 251--268.

\bibitem{Hen2} D.~Hensley, {\em Continued fraction Cantor sets, Hausdorff
dimension, and functional analysis}, J. Number Theory 40 (1992)
336--358.

\bibitem{KerrPinz} D.~Kerr, C.~Pinzari, {\em Noncommutative pressure
and the variational principle in Cuntz--Krieger--type $C^*$-algebras},
J.Funct.Anal. 188 (2002) 156--215.

\bibitem{KLKSS} I.~M.~Khalatnikov, E.~M.~Lifshitz, K.~M.~Khanin, L.~N.~Schur,
Ya.~G.~Sinai. {\it On the stochasticity in relativistic
cosmology.} J.~Stat.~Phys., 38:1/2 (1985), 97--114.


\bibitem{ManMar} Yu.I.~Manin, M.~Marcolli, {\em Continued fractions,
modular symbols, and noncommutative geometry},
Selecta Math. (New Ser.) Vol. 8 N.3 (2002) 475--520.

\bibitem{Mar} M.~Marcolli, {\em Limiting modular symbols and the
Lyapunov spectrum}, J.Number Theory Vol.98 N.2 (2003) 348--376. 

\bibitem{Mayer} D.H.~Mayer, {\em Relaxation properties of the
mixmaster universe},  Phys. Lett. A 121 (1987), no. 8-9,
  390--394. 

\bibitem{Mayer2} D.H.~Mayer, {\em Continued fractions and related
transformations.} In: Ergodic Theory, Symbolic Dynamics
and Hyperbolic Spaces, Eds. T.~Bedford et al., Oxford
University Press, Oxford 1991, pp. 175--222.

\bibitem{MTW} C.W.~Misner, K.S.~Thorne, J.A.~Wheeler, 
{\em Gravitation}, W H Freeman and Co. 1973. 

\bibitem{PoWei} M.~Pollicott, H.~Weiss, {\em Multifractal analysis of
Lyapunov exponent for continued fraction and Manneville-Pomeau
transformations and applications to Diophantine approximation},
Comm. Math. Phys. 207 (1999), no. 1, 145--171.


\end{thebibliography}
\end{document}